\documentclass[aps,prb,twocolumn,showpacs,amsmath,amssymb,superscriptaddress,floatfix]{revtex4-2}
\usepackage[english]{babel}
\usepackage{blindtext}
\usepackage[utf8]{inputenc}
\usepackage{graphicx}
\usepackage{array}
\usepackage{bm}
\usepackage{latexsym}
\usepackage{physics}
\usepackage{amsfonts}
\usepackage{bbold}
\usepackage{scalerel}
\usepackage{appendix}
\let\savediv\div
\let\div\relax
\let\savecorres\corresponds
\let\corresponds\relax
\usepackage{mathabx}
\let\div\savediv
\let\corresponds\savecorres
\usepackage{stmaryrd}
\usepackage[colorlinks = true,
            linkcolor = blue,
            urlcolor  = blue,
            citecolor = blue,
            anchorcolor = blue]{hyperref}
\usepackage{fixltx2e}
\usepackage[usenames,dvipsnames]{color}
\makeatletter
\newcommand*{\rom}[1]{\expandafter\@slowromancap\romannumeral #1@}
\newcommand{\vebm}[1]{
\ifcat\noexpand#1\relax
    {\bm #1}
\else
    {\bf #1}
\fi
}
\newcommand{\posx}{{\vebm{r}}}
\newcommand{\posp}{{\vebm{r}+\vebm{\delta}}}
\newcommand{\kk}{{\vebm{k}}}

\newcommand{\eqrefx}[1]{(\ref{#1})}
\makeatother

\begin{document}

\author{Kirill Alpin}
\affiliation{Max Planck Institute for Solid State Research, Heisenbergstrasse 1, D-70569 Stuttgart, Germany}

%\title{I put a unitary transformation into the Hubbard model and you won't believe what happened next. \emoji{fire}\emoji{ok-hand}\emoji{weary-face}\emoji{weary-face} New strong coupling limit and perturbative description of d-wave superconductivity at finite doping. \emoji{flushed-face}\emoji{flushed-face} Figure\,\ref{fig:YBCOdope} will shock you \emoji{hundred-points}\emoji{hundred-points}}
\title{Perturbative description of high-$T_c$ superconductivity in the Hubbard model via unitary transformation and classical spins}

\date{\today}
\begin{abstract}
%So basically I put a unitary transformation into the hubbard model.
A unitary transformation is applied to the Hubbard model, which maps the Hubbard interaction to a single particle term. The resulting Hamiltonian consists of unconstrained fermions, which is then mapped to a Hamiltonian of spinless fermions coupled to pseudospins. The fermions are integrated out using second order perturbation theory in $1/U$, resulting in an effective spin Hamiltonian. An order parameter is identified, stabilizing d-wave superconductivity. The groundstate energy of classical spin configurations is minimized at a finite value of this order parameter after a critical chemical potential, resulting in d-wave superconductivity at non-zero doping. This suggests, that the onset of high-$T_c$ superconductivity is governed by the groundstate of a classical spin system.

\end{abstract}%
\maketitle

\section{Introduction}
\begin{figure}[t]
    \centering
    \includegraphics[width = 0.45 \textwidth]{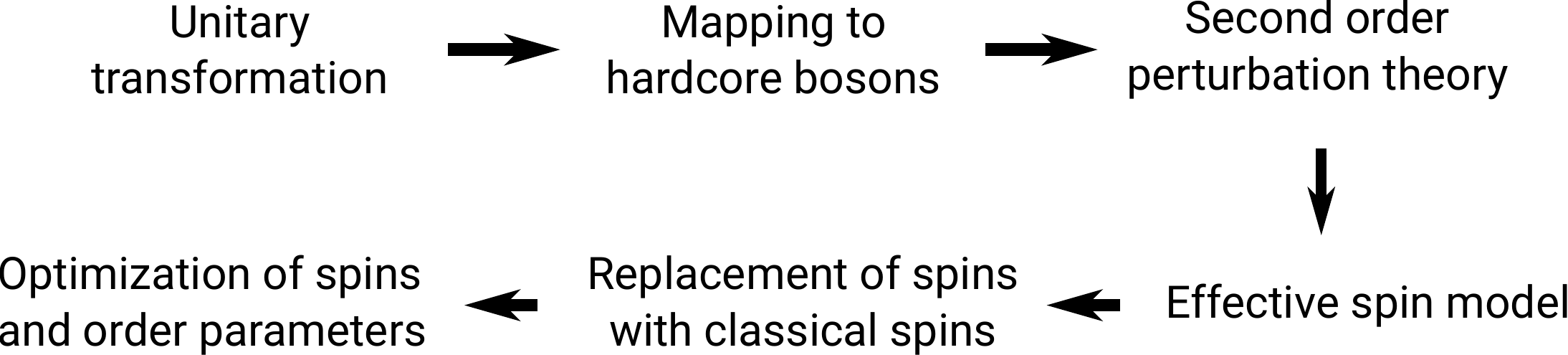}
    \caption{Flow diagram of the $1/U$ perturbative approach described in this paper.}
    \label{fig:flow}
\end{figure}

The Hubbard model in the strong coupling regime is believed to describe the emergence of high-$T_c$ superconductivity in cuprates\,\cite{doi:10.1126/science.235.4793.1196}. The pairing in these compounds is further believed to be d-wave\,\cite{PhysRevLett.73.593}. There have been numerous previous theoretical studies of the Hubbard model\,\cite{qin2022hubbard,doi:10.1146/annurev-conmatphys-031620-102024,RevModPhys.84.1383}. This paper shows a perturbative approach to observing this superconductivity in the strongly repulsive Hubbard model at finite doping. The first step, a unitary transformation, is similar to\,\cite{chowdhury2022sachdev}, where auxiliary spins are coupled to the original Hamiltonian via a unitary Schrieffer-Wolff transformation. Unlike this approach, here the unitary transformation is applied exactly. This results in a Hamiltonian with unconstrained fermions, unlike the t-J model\,\cite{RevModPhys.78.17}. The interaction terms now scale with $t,\mu$ instead of $U$, so perturbative approaches in $1/U$ are viable. This is also not given in the t-J model, where interactions are of the same order as the free part of the Hamiltonian, due to the correlated hopping terms. Therefore, in the t-J model, one must resort to uncontrolled slave-boson approaches\,\cite{RevModPhys.78.17}. The Hamiltonian is then mapped to a system of fermions coupled to pseudospins. In this form, unlike\,\cite{chowdhury2022sachdev}, the Hilbert space is not artificially expanded.
It is solved using variational second order perturbation theory and spins are replaced with classical spins. Beyond\,\cite{PhysRevB.37.9753}, this perturbative expansion can be done at finite doping. An order parameter is identified, which stabilizes d-wave superconductivity. The groundstate energy as a function of this order parameter shows a second order phase transition at a critical chemical potential, resulting in the development of d-wave superconductivity. The flow of the approach described in this paper is shown schematically in Fig.\,\ref{fig:flow}, which we also apply to the Hubbard model with $t'=-0.3t$, showing characteristics of cuprate superconductivity. 

Notable similar approaches are ones based on strong coupling expansions\,\cite{pairault1998strong, pairault2000strong} around the atomic limit of the Hubbard model. Methods like these, in which class for example Dynamical Mean Field Theory (DMFT)\,\cite{PhysRevLett.62.324,PhysRevB.45.6479,RevModPhys.68.13}, Self-Energy Functional Theory\,\cite{potthoff2007self} and Cluster Perturbation Theory\,\cite{potthoff2007self,senechal2002cluster} belong to, capture the local correlations of a given system, while it is a challenge to incorporate non-local self-energies\,\cite{RevModPhys.77.1027,RevModPhys.90.025003}. The unitary transformation defined in this manuscript is applied exactly, so no correlations are lost there. The perturbative expansion following can in principle capture any non-locality, since no diagrams are discarded and this series can be carried out to higher order than the one in this paper, as it has been done in studies using DiagMC\,\cite{PhysRevB.104.L020507}. Although there, the expansion is carried out in the weak coupling regime, while here we offer to do the same at strong coupling (using for example ordinary diagrammatics in the case of the Hamiltonian with unconstrained fermions). Other weak coupling expansions include\,\cite{PhysRevB.94.085106,PhysRevB.67.035112}.
\section{Local unitary transformation}
The square lattice repulsive Hubbard model with $U>0$ and $t<0$ is given by
\begin{align}
    H&=\sum_{\vebm{r},\vebm{\delta},\sigma} t (c^\dag_{\vebm{r},\sigma}c_{\vebm{r}+\vebm{\delta},\sigma}
    +c^\dag_{\vebm{r}+\vebm{\delta},\sigma}c_{\vebm{r},\sigma}
    )\nonumber\\ &+\sum_{\vebm{r}} U \left(n_{\posx,\uparrow} - \frac{1}{2}\right)\left(n_{\posx,\downarrow} - \frac{1}{2}\right)+\sum_{\vebm{r},\sigma} \mu n_{\posx,\sigma}. \label{H}
\end{align}
In this convention half filling is at $\mu=0$ and the system is hole doped with $\mu>0$. $\vebm{\delta}$ runs over the two vectors $(1,0)$, $(0,1)$. We apply a unitary transformation
\begin{align}
    \mathcal{U}=\prod_{\vebm{r}} U_\vebm{r}
\end{align}
to the Hamiltonian (\ref{H}), which decomposes into a series of local unitaries $U_\vebm{r}$. We consider the following unitary
\begin{align}
    U_\vebm{r} = 1 - n_{\vebm{r},2} - n_{\vebm{r},2} \Bar{c}_{\vebm{r},1} \Tilde{c}_{\vebm{r},3} \label{U}
\end{align}
where we relabeled $c_{\vebm{r},1} \equiv c_{\vebm{r},\uparrow}$, $c_{\vebm{r},2} \equiv c_{\vebm{r},\downarrow}$. We have introduced a third fermion species $c_{\vebm{r},3}$ at every lattice site, which has no contribution in the original Hamiltonian \eqrefx{H}. It does not contribute to any observables. This auxiliary fermion is necessary to construct a parity conserving unitary operation. Parity conservation is needed to form a local $U_\posx$ (a non-parity conserving version of this transformation, dropping the third fermion species, can still be used in 1D).

We used the following definitions in Eq.\,\eqrefx{U} 
\begin{align}
    \Bar{c}_{\vebm{r},n}&\equiv c^\dag_{\vebm{r},n}-c_{\vebm{r},n},\\
    \Tilde{c}_{\vebm{r},n}&\equiv c^\dag_{\vebm{r},n}+c_{\vebm{r},n}.
\end{align}
The unitary transformation\,\eqrefx{U} performs the following mapping from the original basis $\ket{\text{fermions},n_3}$ to the $\ket{n_1,n_2,n_3}$ basis
\begin{align}
\ket{0,0} &\shortrightarrow \ket{0,0,0}, \quad
\ket{\uparrow,0} \shortrightarrow \ket{1,0,0}, \\
\ket{\downarrow,0} &\shortrightarrow \ket{1,1,1}, \quad
\ket{\updownarrows,0} \shortrightarrow \ket{0,1,1}, \\
\ket{0,1} &\shortrightarrow \ket{0,0,1}, \quad
\ket{\uparrow,1} \shortrightarrow \ket{1,0,1}, \\
\ket{\downarrow,1} &\shortrightarrow \ket{1,1,0}, \quad
\ket{\updownarrows,1} \shortrightarrow \ket{0,1,0}.
\end{align}
Notice, that all, at large $U$ and half filling, prohibited states, so double and $0$ occupancy states, are mapped to a vacuum state of fermion species 1, while all low energy states are mapped to ones, where fermion 1 is fully occupied. This allows us, to effectively integrate out double and $0$ occupancies by perturbation theory.
With 
\begin{align}
    \Bar{c}^2_{\vebm{r},n}=-1, \quad \Tilde{c}^2_{\vebm{r},n}=1
\end{align}
we can show that $U^\dag_\vebm{r}=U_\vebm{r}$ and that $U^\dag_\vebm{r}U_\vebm{r}=1=U^2_\vebm{r}$. So $\mathcal{U}$ is unitary $\mathcal{U}^\dag \mathcal{U}=1$. Because $U_\vebm{r}$ is parity conserving, it commutes with itself at any other site (and on the same site) $[U_\vebm{r},U_\vebm{r'}]=0$.

Applying $U_\vebm{r}$ to the Hubbard interaction maps it to a single particle term
\begin{align}
    U_\vebm{r} \left(n_{\uparrow} - \frac{1}{2}\right)\left(n_{\downarrow} - \frac{1}{2}\right) U_\vebm{r} = \frac{1}{4} - \frac{1}{2} n_{\vebm{r},1}
\end{align}
where we used $U^\dag_\vebm{r}=U_\vebm{r}$.
The chemical potential term is mapped to 
\begin{align}
    U_\vebm{r} \sum_{\sigma} n_{\vebm{r},\sigma} U_\vebm{r} = n_{\vebm{r},1} + 2 n_{\vebm{r},2} (1 - n_{\vebm{r},1}).
\end{align}
Finally, the hopping term is mapped to
\begin{align}
    &U_{\vebm{r}+\vebm{\delta}} U_\vebm{r} \sum_{\sigma}(c^\dag_{\vebm{r},\sigma}c_{\vebm{r}+\vebm{\delta},\sigma}
    +c^\dag_{\vebm{r}+\vebm{\delta},\sigma}c_{\vebm{r},\sigma}
    ) U_\vebm{r} U_{\vebm{r}+\vebm{\delta}} \\
    =&K_{\vebm{r},\vebm{\delta}}c^\dag_{\vebm{r},1}c^\dag_{\vebm{r}+\vebm{\delta},1} + (1+A_{\vebm{r},\vebm{\delta}})c^\dag_{\vebm{r},1}c_{\vebm{r}+\vebm{\delta},1} + \text{h.c.}
\end{align}
with
\begin{align}
    K_{\vebm{r},\vebm{\delta}}=&-n_{\vebm{r},2}+n_{\vebm{r}+\vebm{\delta},2}\nonumber\\&
    -(c^\dag_{\vebm{r},2}c_{\vebm{r}+\vebm{\delta},2} + c^\dag_{\vebm{r}+\vebm{\delta},2}c_{\vebm{r},2})\Tilde{c}_{\vebm{r},3}\Tilde{c}_{\vebm{r}+\vebm{\delta},3},\\
    A_{\vebm{r},\vebm{\delta}}=&-n_{\vebm{r},2}-n_{\vebm{r}+\vebm{\delta},2}\nonumber\\&+(c^\dag_{\vebm{r},2}c_{\vebm{r}+\vebm{\delta},2} + c^\dag_{\vebm{r}+\vebm{\delta},2}c_{\vebm{r},2})\Tilde{c}_{\vebm{r},3}\Tilde{c}_{\vebm{r}+\vebm{\delta},3}.
\end{align}
Let's apply $\mathcal{U}$ to the whole Hubbard model\,\eqrefx{H}
\begin{align}
    \mathcal{U}^\dag H \mathcal{U} &= \prod_{\vebm{r}'} U_{\vebm{r}'} H \prod_{\vebm{r}} U_\vebm{r}\\
    &= t \sum_{\vebm{r},\vebm{\delta},\sigma} U_{\vebm{r}+\vebm{\delta}} U_\vebm{r} (c^\dag_{\vebm{r},\sigma}c_{\vebm{r}+\vebm{\delta},\sigma}
    +c^\dag_{\vebm{r}+\vebm{\delta},\sigma}c_{\vebm{r},\sigma}
    ) U_\vebm{r} U_{\vebm{r}+\vebm{\delta}}
    \nonumber\\ &+U \sum_{\vebm{r}} U_\vebm{r} \left(n_{\posx,\uparrow} - \frac{1}{2}\right)\left(n_{\posx,\downarrow} - \frac{1}{2}\right) U_\vebm{r}\nonumber\\ 
    &+\mu \sum_{\vebm{r},\sigma} U_\vebm{r} n_{\posx,\sigma} U_\vebm{r}.
\end{align}
The order of the application of $U_\posx$ in the two products above is not important, since $U_\posx$ commute. We further used that for any bosonic or fermionic operators $X_\vebm{a}, Y_{\vebm{b}}$ we have
\begin{align}
    &\prod_{\vebm{r}'} U_{\vebm{r}'} X_\vebm{a} Y_{\vebm{b}} \prod_{\vebm{r}} U_{\vebm{r}} \\
    &= U_{\vebm{b}} U_\vebm{a} X_\vebm{a} Y_{\vebm{b}} U_\vebm{a} U_{\vebm{b}}
\end{align}
because $[X_\vebm{a} Y_{\vebm{b}}, U_\vebm{r}]=0$ as long as $\vebm{a}\neq\vebm{r}$ and $\vebm{b}\neq\vebm{r}$.

So the transformed Hamiltonian becomes
\begin{align}
    H' &= \mathcal{U}^\dag H \mathcal{U} \nonumber\\
    &= t \sum_{\vebm{r},\vebm{\delta}} 
    K_{\vebm{r},\vebm{\delta}}c^\dag_{\vebm{r},1}c^\dag_{\vebm{r}+\vebm{\delta},1} + (1+A_{\vebm{r},\vebm{\delta}})c^\dag_{\vebm{r},1}c_{\vebm{r}+\vebm{\delta},1} + \text{h.c.}
    \nonumber\\ &+ \sum_{\vebm{r}} -\frac{U}{2} n_{\vebm{r},1} + \mu (n_{\vebm{r},1} + 2 n_{\vebm{r},2} (1 - n_{\vebm{r},1})). \label{Hp}
\end{align}
Since the Hubbard interaction is mapped to just a chemical potential of the fermion species 1, the interactions in $K_{\vebm{r},\vebm{\delta}}$ and $A_{\vebm{r},\vebm{\delta}}$ now scale with $t$ instead of $U$. Also notice, that, unlike the t-J model, Hamiltonian \eqrefx{Hp} describes {\bf unconstrained} fermions. Therefore, in this basis, the large interaction limit $U\gg t$ can be studied perturbatively at $\mu\neq 0$. Further, the mapping we have done is {\bf exact} and works at any $U,t,\mu$.

Note that $H'$ commutes with any bond operator $i\Bar{c}_{\vebm{r},3}\Bar{c}_{\vebm{r}',3}$, so one could block diagonalize it in the eigenbasis of all $i\Bar{c}_{\vebm{r},3}\Bar{c}_{\vebm{r}',3}$.

Ultimately, we would like to compute observables in this Hamiltonian. To do so, we also need to transform those too. The transformed magnetization is
\begin{align}
    U_\vebm{r} (n_{\vebm{r},\uparrow}-n_{\vebm{r},\downarrow}) U_\vebm{r} = n_{\vebm{r},1} (1 - 2 n_{\vebm{r},2}).
\end{align}
The transformed particle density is
\begin{align}
    U_\vebm{r} (n_{\vebm{r},\uparrow}+n_{\vebm{r},\downarrow}) U_\vebm{r} = 2 n_{\vebm{r},2} (1 - n_{\vebm{r},1}) + n_{\vebm{r},1}.\label{particledens}
\end{align}
A two site singlet pair operator transformed into the new basis is
\begin{align}
    &\Delta = U_{\vebm{r}+\vebm{\delta}} U_\vebm{r} \sum_\sigma (-1)^\sigma ( c_{\vebm{r}+\vebm{\delta},\Bar{\sigma}} c_{\vebm{r},\sigma} ) U_\vebm{r} U_{\vebm{r}+\vebm{\delta}}=\\&
    c_{\posx,1}c_{\posp,1} (-c_{\posx,2}\Tilde{c}_{\posx,3}+c_{\posp,2}\Tilde{c}_{\posp,3}\nonumber\\&
    -n_{\posx,2}c_{\posp,2}\Tilde{c}_{\posp,3}+n_{\posp,2}c_{\posx,2}\Tilde{c}_{\posx,3}) +\nonumber\\&
    c^\dag_{\posp,1}c_{\posx,1} ( c_{\posx,2}\Tilde{c}_{\posx,3} + n_{\posx,2}c_{\posp,2}\Tilde{c}_{\posp,3} - n_{\posp,2}c_{\posx,2}\Tilde{c}_{\posx,3}) +\nonumber\\&
    c^\dag_{\posx,1}c_{\posp,1} ( c_{\posp,2}\Tilde{c}_{\posp,3} - n_{\posx,2}c_{\posp,2}\Tilde{c}_{\posp,3} + n_{\posp,2}c_{\posx,2}\Tilde{c}_{\posx,3})+ \nonumber\\&
    c^\dag_{\posp,1}c^\dag_{\posx,1} (n_{\posx,2}c_{\posp,2}\Tilde{c}_{\posp,3} - n_{\posp,2}c_{\posx,2}\Tilde{c}_{\posx,3})\label{origdelta}
\end{align}
with $\Bar{\sigma}=-\sigma$, $\uparrow=0$ and $\downarrow=1$.

\section{Perturbation theory}
In the groundstate and at large $U$ and $\mu=0$, the fermion species 1 is, due to the $-\frac{U}{2}$ chemical potential, fully occupied. In this limit, we can therefore perform perturbation theory around $\ket{\Psi}=\prod_\posx c^\dag_{\posx,1} \ket{0}$. 

The first order is zero. Second order is
\begin{align}
    H_2 = -\frac{1}{U} \bra{\Psi} H'^2 \ket{\Psi}.
\end{align}
The factor $-\frac{1}{U}$ comes from the energy needed to excite two $c_1$ fermions. $H_2$ becomes
\begin{align}
H_2 = &-\frac{2t^2}{U}\sum_{\vebm{r},\vebm{\delta}} n_{\posx,2} + n_{\posp,2} - 2 n_{\posx,2}n_{\posp,2} \nonumber\\
&-(c^\dag_{\posx,2}c_{\posp,2}-c^\dag_{\posp,2}c_{\posx,2})\Tilde{c}_{\posx,3}\Tilde{c}_{\posp,3}.
\end{align}
Consider the following composite particles
\begin{align}
    b^\dag_\posx = \Tilde{c}_{\posx,3}c^\dag_{\posx,2} \quad b_{\posx}=c_{\posx,2}\Tilde{c}_{\posx,3}.\label{dronefermions}
\end{align}
We find
\begin{align}
    [b_\posx,b_{\vebm{r}'}]&=0\\
    [b^\dag_\posx,b^\dag_{\vebm{r}'}]&=0\\
    [b_\posx,b^\dag_{\vebm{r}'}]&=(1-2b^\dag_\posx b_{\posx} )\delta_{\posx,\vebm{r}'}.
\end{align}
These are the commutation relations of hardcore bosons. Eq.\,\eqrefx{dronefermions} is equivalent to the drone-fermion representation of spins\,\cite{khaliullin1994spin}. $n_{\posx,2}$ can be written in $b$ like so
\begin{align}
    n_{\posx,2} = &c^\dag_{\posx,2}c_{\posx,2} = 
    c^\dag_{\posx,2}c_{\posx,2}\Tilde{c}^2_{\posx,3} \\= &\Tilde{c}_{\posx,3}c^\dag_{\posx,2}c_{\posx,2}\Tilde{c}_{\posx,3} 
    = b^\dag_\posx b_\posx = n_{\posx,b}
\end{align}
where we used $\Tilde{c}^2_{\posx,3}=1$. $H_2$ in hardcore bosons is
\begin{align}
H_2 = &-\frac{2t^2}{U}\sum_{\vebm{r},\vebm{\delta}} n_{\posx,b} + n_{\posp,b} - 2 n_{\posx,b}n_{\posp,b} \nonumber\\
&-b^\dag_{\posx,2}b_{\posp,2}-b^\dag_{\posp,2}b_{\posx,2}.
\end{align}
Further, we can write hardcore bosons as spins
\begin{align}
    S^+_\posx =&b_\posx \quad S^-_\posx=b^\dag_\posx \label{pseudospins}\\
    S^z_\posx = &\frac{1}{2} - b^\dag_\posx b_\posx \\
    S^x_\posx = &\frac{1}{2} (b_\posx + b^\dag_\posx) \\
    S^y_\posx = &\frac{1}{2i} (b_\posx - b^\dag_\posx)
\end{align}
This reproduces the well known $U\gg1$ limit of the Hubbard model at half filling, that is the Heisenberg model
\begin{align}
H_2 = &\frac{4t^2}{U}\sum_{\vebm{r},\vebm{\delta}} \vebm{S}_\posx \cdot \vebm{S}_\posp.\label{Heisenberg}
\end{align}
Notice that the reverse transformation of a general spin Hamiltonian, that is first to hardcore bosons and then to the composite fermions containing $\Tilde{c}_{\posx,3}$ can always be done and results in a Hamiltonian with unconstrained fermions\,\cite{khaliullin1994spin}.

\section{Variational perturbation theory at finite doping}
In this section, we perform perturbation theory at finite $\mu$ with the goal of calculating the expectation value of $\Delta$. In terms of hardcore bosons, the $\Delta$ order parameter in Eq.\,\eqrefx{origdelta} transforms to
\begin{align}
    &\Delta = 
    c_{\posx}c_{\posp} (-b_\posx+b_\posp\nonumber\\&
    -n_{\posx,b}b_\posp+n_{\posp,b}b_\posx) +\nonumber\\&
    c^\dag_{\posp}c_{\posx} ( b_\posx + n_{\posx,b}b_\posp - n_{\posp,b}b_\posx) +\nonumber\\&
    c^\dag_{\posx}c_{\posp} ( b_\posp - n_{\posx,b}b_\posp + n_{\posp,b}b_\posx) +\nonumber\\&
    c^\dag_{\posp}c^\dag_{\posx} (n_{\posx,b}b_\posp - n_{\posp,b}b_\posx).
\end{align}
Hardcore bosons can be expressed as spin operators, which in turn reduces the singlet order parameter $\Delta$ to a spin wave order parameter. Note, that an uneven number of hardcore boson operators are present in $\Delta$. When the expectation value of $\Delta$ is computed, even at first order in perturbation, an uneven number of hardcore boson operators remain. Translated into the spin picture, this means that to have a finite expectation value of $\Delta$, that is superconductivity in any way, pseudospin U(1) symmetry must be broken. This U(1) symmetry breaking can also be seen when $\Delta$ is multiplied with a global U(1) gauge of the original fermionic operators $c_{\posx,\sigma}\rightarrow\exp(i\theta/2)c_{\posx,\sigma}$. The only place where this phase can consistently be absorbed to are the $b_\posx$ operators. A phase in $\exp(i\theta)b_\posx$ translates to rotations around the z-axis of the pseudospins $\cos(\theta)S_x+\sin(\theta)S_y$, see Eq.\,\eqrefx{pseudospins}. So charge $U(1)$ symmetry is transformed using the local unitary transformation\,\eqrefx{U} to a pseudospin $U(1)$ symmetry, which we expect to be broken in a superconductor. We can transform the Hamiltonian \eqrefx{Hp} into one containing hardcore bosons
\begin{align}
    H' 
    &= t \sum_{\vebm{r},\vebm{\delta}} 
    K_{\vebm{r},\vebm{\delta}}c^\dag_{\vebm{r}}c^\dag_{\vebm{r}+\vebm{\delta}} + (1+A_{\vebm{r},\vebm{\delta}})c^\dag_{\vebm{r}}c_{\vebm{r}+\vebm{\delta}} + \text{h.c.}
    \nonumber\\ &+ \sum_{\vebm{r}} -\frac{U}{2} n_{\vebm{r}} + \mu (n_{\vebm{r}} + 2 n_{\vebm{r},b} (1 - n_{\vebm{r}})). \label{Hf}
\end{align}
We dropped the 1 index for the $c_1$ fermions. The $K$ and $A$ terms expressed in hardcore bosons are
\begin{align}
    K_{\posx,\vebm{\delta}} &= -n_{\posx,b} + n_{\posp,b} - b^\dag_\posx b_\posp +b^\dag_\posp b_\posx \\
    A_{\posx,\vebm{\delta}} &= -n_{\posx,b} - n_{\posp,b} + b^\dag_\posx b_\posp -b^\dag_\posp b_\posx 
\end{align}
Fourier transforming everything, we end up with the following interaction vertices
\begin{align}
    V_K & = \sum_{\vebm{\delta}}t K_{\vebm{r},\vebm{\delta}}c^\dag_{\vebm{r}}c^\dag_{\vebm{r}+\vebm{\delta}} + \text{h.c.}\nonumber\\
    &= \sum_{\vebm{\delta}}t b^\dag_{\kk_1} b_{\kk_2} \vebm{c}^\dag_{\kk_3} \cdot (-k(\vebm{0},\vebm{0},\vebm{\delta}) + k(\vebm{\delta},\vebm{\delta},\vebm{\delta}) \nonumber\\
    &-k(\vebm{0},\vebm{\delta},\vebm{\delta}) + k(\vebm{\delta},\vebm{0},\vebm{\delta})
    ) \cdot \vebm{c}_{\kk_4}
\end{align}
\begin{align}
    V_{A+\mu} &= -2\mu n_{\posx,b} n_{\posx}+\sum_{\vebm{\delta}}t(A_{\vebm{r},\vebm{\delta}}c^\dag_{\vebm{r}}c_{\vebm{r}+\vebm{\delta}} +
    A^\dag_{\vebm{r},\vebm{\delta}}c^\dag_{\vebm{r}+\vebm{\delta}}c_{\vebm{r}})
    \nonumber\\
    &= b^\dag_{\kk_1} b_{\kk_2} \vebm{c}^\dag_{\kk_3} \cdot 
    (-2\mu a(\vebm{0},\vebm{0},\vebm{0}) \nonumber\\&+ t\sum_{\vebm{\delta}} a(\vebm{0},\vebm{0},\vebm{\delta}) - a(\vebm{\delta},\vebm{\delta},\vebm{\delta}) \nonumber\\
    &+a(\vebm{0},\vebm{\delta},\vebm{\delta}) - a(\vebm{\delta},\vebm{0},\vebm{\delta})
    ) \cdot \vebm{c}_{\kk_4}
\end{align}
with 
\begin{align}
k(\vebm{\delta}_1,\vebm{\delta}_2,\vebm{\delta})&= \begin{pmatrix}
0 & e^{i\kk_4 \vebm{\delta}} e^{-i \kk_1 \vebm{\delta}_1 + i \kk_2 \vebm{\delta}_2} \\
e^{-i \kk_3 \vebm{\delta}} e^{-i \kk_1 \vebm{\delta}_2 + i \kk_2 \vebm{\delta}_1} & 0
\end{pmatrix} \\
    a(\vebm{\delta}_1,\vebm{\delta}_2,\vebm{\delta})&= \begin{pmatrix}
    e^{i\kk_4 \vebm{\delta} -i \kk_1 \vebm{\delta}_1 + i \kk_2 \vebm{\delta}_2} + e^{-i \kk_3 \vebm{\delta} -i \kk_1 \vebm{\delta}_2 + i \kk_2 \vebm{\delta}_1} & 0 \\
    0 & 0
    \end{pmatrix}
\end{align}
\begin{align}
    \vebm{c}_{\kk} &= \begin{pmatrix} c_{\kk} \\ c^\dag_{-\kk} \end{pmatrix}
\end{align}
We dropped the sum over $\posx$ and $\kk_n$ and the momentum conserving $\delta(\kk_a+\kk_b+...)$ in all equations above. We see, that the chemical potential term acts as a magnetic field for the pseudospins, making it possible to break $U(1)$ symmetry of the pseudospins, as discussed before.

We continue to perform second order perturbation theory on this Hamiltonian at finite $\mu$, to derive an effective hardcore boson and therefore pseudospin Hamiltonian. We note here, that there is in general an arbitrariness in how one can perform perturbation theory. Usually, perturbations around $H_0$ only contain the interaction terms $V$ of the Hamiltonian $H=H_0+V$. In principle it is possible to add and subract a non-interacting term $V_\lambda$ like so $H=H_0-V_\lambda+V+V_\lambda$. One can define a new $H_0'=H_0-V_\lambda$ and perturbation around it $V'=V+V_\lambda$. This way it is possible to describe symmetry broken phases using a perturbative expansion, as long as $V_\lambda$ is much smaller than $H_0$. The right $V_\lambda$ is chosen variationally. A similar perturbative approach has been applied directly to the Hubbard model at weak coupling in\,\cite{PhysRevB.67.035112}, where it was possible to optimize the introduced fields via a self-consistency equation and where a finite d-wave superconductive field has been found at non-zero doping. 

One of the terms we vary between $H_0$ and $V_\lambda$ is the hopping term $t c^\dag_\posx c_\posp + \text{h.c.}$ Without variational perturbation theory (VPT), it is unclear if one has to put this term in $H_0$ or $V'$, as it is of order $t$ while still being a free particle term.

Before going on to the results of VPT, we describe in the following the way VPT is implemented. We first define $V_{\lambda=0}=0$, and $H_0'=H_0(\lambda)$ such that $H_0(\lambda=0)=H_0$. The zero and first order terms are
\begin{align}
    H_f = \bra{\Psi(\lambda)}H_0(\lambda=0) + V_K + V_{A+\mu}\ket{\Psi(\lambda)}
\end{align}
where $\ket{\Psi(\lambda)}$ is the groundstate of $H_0(\lambda)$. In second order we must compute
\begin{align}
    &\bra{\Psi(\lambda)} V \frac{1}{E_0(\lambda) - H_0(\lambda)} V \ket{\Psi(\lambda)}
\end{align}
where $V=V_K + V_{A+\mu} + V_{\lambda}$ and where we only consider connected diagrams, such that we can drop the projection operators to excited states. Due to the structure of all of the present vertices, the computation of the second order terms reduce to the evaluation of
\begin{align}
    &\frac{1}{E_0(\lambda) - H_0(\lambda)}(\vebm{A}\cdot\vebm{c}^\dag_{\kk})(\vebm{B}\cdot\vebm{c}_{\kk'})\ket{\Psi(\lambda)}\nonumber\\
    =&\frac{1}{E_0(\lambda) - H_0(\lambda)}(\vebm{A}\sigma_x\cdot\vebm{c}_{-\kk})(\vebm{B}\cdot\vebm{c}_{\kk'})\ket{\Psi(\lambda)}
\end{align}
with some vectors $\vebm{A}$ and $\vebm{B}$ and 
where we used that 
\begin{align}
    \vebm{c}^\dag_{\kk} &= \sigma_x \vebm{c}_{-\kk}
\end{align}
is the particle-hole transformation in the Nambu formalism.
With the commutation relation
\begin{align}
    [H_0,(\vebm{A}\sigma_x\cdot\vebm{c}_{-\kk})(\vebm{B}\cdot\vebm{c}_{\kk'})] &= (\sigma_x H_{\kk} \sigma_x\vebm{A}\sigma_x \cdot\vebm{c}_{-\kk})(\vebm{B}\cdot\vebm{c}_{\kk'}) \nonumber\\&+
    (\vebm{A}\sigma_x \cdot\vebm{c}_{-\kk})(\sigma_x H_{-\kk'}\sigma_x \vebm{B}\cdot\vebm{c}_{\kk'})
\end{align}
where $H_{\kk}$ is defined by $H_0$
\begin{align}
    H_0(\lambda) = \frac{1}{2}\sum_{\kk} \vebm{c}^\dag_{\kk} H_{\kk}(\lambda) \vebm{c}_{\kk},
\end{align}
we can show
\begin{align}
    &(1 - E_0 + H_0) (\vebm{A}\sigma_x\cdot\vebm{c}_{-\kk})(\vebm{B}\cdot\vebm{c}_{\kk'})\ket{\Psi(\lambda)} = \nonumber\\&
    (\sigma_x H_{\kk}\sigma_x \vebm{A}\sigma_x \cdot\vebm{c}_{-\kk})(\vebm{B}\cdot\vebm{c}_{\kk'}) \nonumber\\&+
    (\vebm{A}\sigma_x \cdot\vebm{c}_{-\kk})(\sigma_x H_{-\kk'}\sigma_x\vebm{B}\cdot\vebm{c}_{\kk'}) \nonumber\\&+ (\vebm{A}\sigma_x\cdot\vebm{c}_{-\kk})(\vebm{B}\cdot\vebm{c}_{\kk'})\ket{\Psi(\lambda)}.
\end{align}
We can define a two particle space
\begin{align}
    (\vebm{A}\sigma_x\otimes\vebm{B})(\vebm{c}_{-\kk}\otimes\vebm{c}_{\kk'})\ket{\Psi_1}.
\end{align}
Acting with $1-E_0+H_0$ on this state gives
\begin{align}
&(1+\sigma_xH_{\kk}\sigma_x\otimes\mathbb{1} + \mathbb{1} \otimes \sigma_x H_{-\kk'}\sigma_x)\nonumber\\&\vebm{A}\sigma_x\otimes\vebm{B}\cdot\vebm{c}_{-\kk}\otimes\vebm{c}_{\kk'}\ket{\Psi(\lambda))}.
\end{align}
So 
\begin{align}
    &\frac{1}{E_0-H_0}\vebm{A}\sigma_x\otimes\vebm{B}\cdot\vebm{c}_{-\kk}\otimes\vebm{c}_{\kk'}\ket{\Psi(\lambda)}\nonumber\\
    =&\sum_{n=0}^\infty (1 - E_0 + H_0)^n \vebm{A}\sigma_x\otimes\vebm{B}\cdot\vebm{c}_{-\kk}\otimes\vebm{c}_{\kk'}\ket{\Psi(\lambda)}\nonumber\\
    =&\frac{1}{- \sigma_xH_{\kk}\sigma_x\otimes\mathbb{1} - \mathbb{1} \otimes \sigma_xH_{-\kk'}\sigma_x} \vebm{A}\sigma_x\otimes\vebm{B}\cdot\vebm{c}_{-\kk}\otimes\vebm{c}_{\kk'}\ket{\Psi(\lambda)}.
\end{align}
$\sigma_xH_{\kk}\sigma_x\otimes\mathbb{1} + \mathbb{1} \otimes \sigma_xH_{-\kk'}\sigma_x$ can be seen as the free two particle Hamiltonian propagating two holes. In general, vertices have a $\kk,\kk'$ dependence
\begin{align}
    \vebm{A}\sigma_x\otimes\vebm{B}\cdot\vebm{c}_{-\kk}\otimes\vebm{c}_{\kk'} = \vebm{c}_{-\kk} (\sigma_x V(\kk,\kk')) \vebm{c}_{\kk'}.
\end{align}
So in practice, to apply $\frac{1}{E_0-H_0}$ to this vertex, $- \sigma_xH_{\kk}\sigma_x\otimes\mathbb{1} - \mathbb{1} \otimes \sigma_xH_{-\kk'}\sigma_x$ is first diagonalized and $\sigma_x V(\kk,\kk')$ is transformed into a basis where the two particle propagator is diagonal. Then it is straightforward to apply $\frac{1}{E_0-H_0}$ and the transformation is reverted. Afterwards we arrive at
\begin{align}
    &\frac{1}{E_0-H_0}\vebm{c}_{-\kk} (\sigma_x V(\kk,\kk')) \vebm{c}_{\kk'}\ket{\Psi(\lambda)}\nonumber\\
    =&\vebm{c}_{-\kk} (\sigma_x \tilde{V}(\kk,\kk')) \vebm{c}_{\kk'}\ket{\Psi(\lambda)}.
\end{align}
Then we only need to contract expectation values of the following form
\begin{align}
    \bra{\Psi(\lambda)}\vebm{c}^\dag_{\kk_1} (V(\kk_1,\kk_2)\sigma_x) \vebm{c}^\dag_{-\kk_2}\vebm{c}_{-\kk_3} (\sigma_x \tilde{V}(\kk_3,\kk_4)) \vebm{c}_{\kk_4}\ket{\Psi(\lambda)}
\end{align}
which can be done using Wick's theorem. Leaving only connected diagrams, this results with Einstein summation in
\begin{align}
&(V(\kk_4,\kk_3)\sigma_x)_{ab}(\sigma_x \tilde{V}(\kk_3,\kk_4))_{cd}\nonumber\\&\rho(\kk_4)^{ad}\delta_{k_1,k_4}\rho(-\kk_3)^{bc}\delta_{k_2,k_3}\nonumber\\
-&(V(\kk_3,\kk_4)\sigma_x)_{ab}(\sigma_x \tilde{V}(\kk_3,\kk_4))_{cd}\nonumber\\&\rho(-\kk_3)^{ac}\delta_{k_1,-k_3}\rho(\kk_4)^{bd}\delta_{-k_2,k_4}
\end{align}
where $\rho(\kk)=\langle\vebm{c}^\dag_{\kk}\vebm{c}_{\kk}\rangle$ are the one particle density matrices of $\frac{1}{2}H_\kk$. In the following, $\rho$ is computed at a temperature of $T=10^{-4}$. Note that for interacting vertices $V$ contains $b$ operators, resulting in an effective hardcore boson Hamiltonian, with terms of the form
\begin{align}
    \sum_\kk \epsilon_f(\kk) b^\dag_\kk b_\kk
\end{align}
and interactions
\begin{align}
    \sum_\kk v_b(\kk,\kk',\vebm{q}) b^\dag_\kk b_{\kk+\vebm{q}} b^\dag_{\kk'} b_{\kk'-\vebm{q}}.
\end{align}
Both terms are Fourier transformed back to real space, after which hardcore bosonic commutation relations are applied to result in the simplest form of a given expression. For example
\begin{align}
    b^\dag_{\posx_1} b_{\posx_2} b^\dag_{\posx_1} b_{\posx_3} = 0
\end{align}
and
\begin{align}
    b^\dag_{\posx_1} b_{\posx_2} b^\dag_{\posx_2} b_{\posx_2} = b^\dag_{\posx_1} b_{\posx_2}.
\end{align}
The resulting Hamiltonian is then transformed to a pseudospin Hamiltonian using Eq.\,\eqrefx{pseudospins}. To compute the groundstate energy and expectation values, spins are replaced by classical spins of length $1/2$. The spins are chosen to minimize the classical groundstate energy. This approximation is the first step of spin wave theory and is valid in an ordered phase with low densities of excitations around the classical groundstate. All of the above steps are done algorithmically. Integrations over the BZ are done on a regular grid and the minimization of the classical spins energy is implemented using the Pytorch library\,\cite{NEURIPS2019_9015} and an Adam optimizer\,\cite{kingma2014adam}.

\subsection{Pure Hubbard model at $U=8$}
\begin{figure}[t]
    \centering
    \includegraphics[width = 0.5 \textwidth]{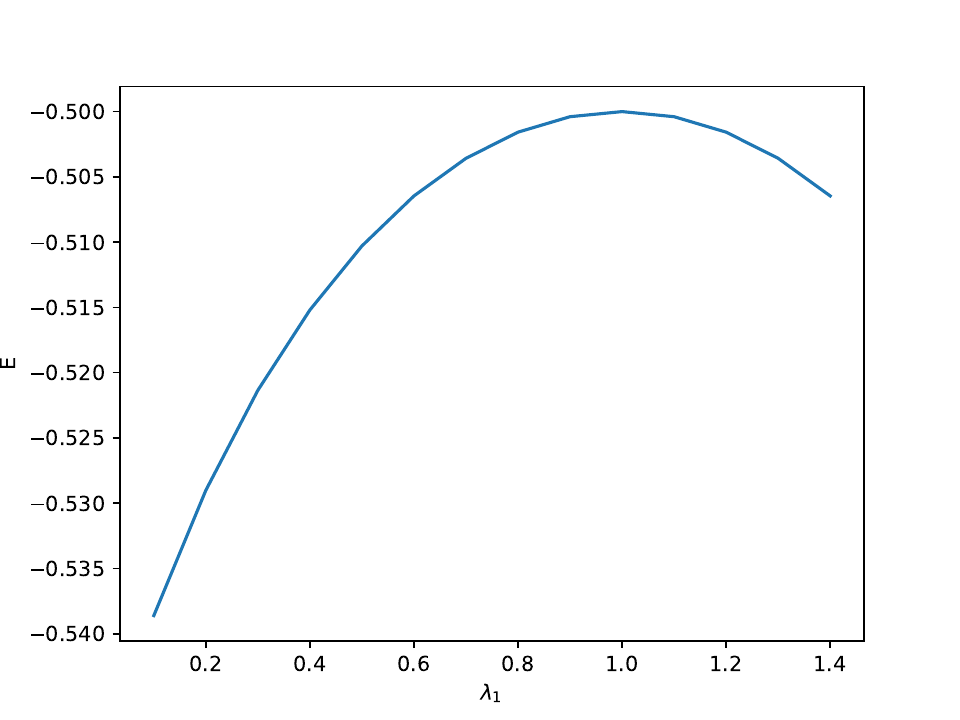}
    \caption{Groundstate energy over $\lambda_1$ for $\mu=0$ and $U=8$, which shows an optimum at $\lambda_1=1$. Integrations for interaction vertices were done on a 16x16 grid BZ. For all other integrations a 64x64 grid was used.}
    \label{fig:sweept0}
\end{figure}

\begin{figure}[t]
    \centering
    \includegraphics[width = 0.5 \textwidth]{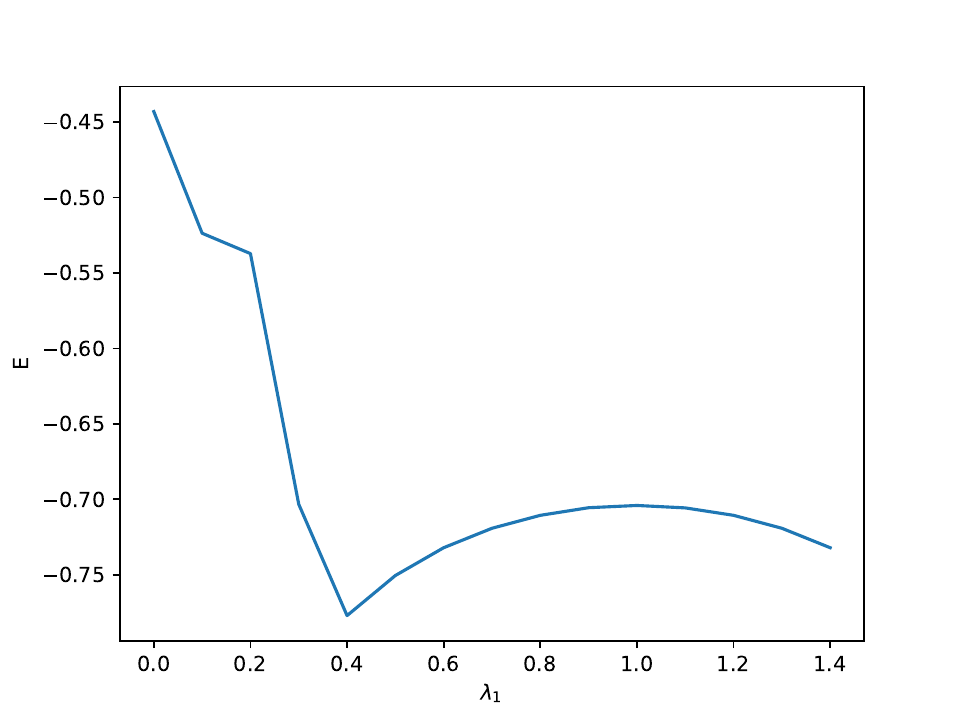}
    \caption{Groundstate energy over $\lambda_1$ for $\mu=1.5$ and $U=8$. The optimum adiabatically connected to the AFM groundstate at $\mu=0$ persists. At $\lambda_1=0.4$ a phase transition to a long range pseudospin stripe groundstate appears, where no optimum can be found. The resulting minimum at the phase transition is discontinuous and therefore not stable. Integrations for interaction vertices were done on a 16x16 grid BZ. For all other integrations a 64x64 grid was used.}
    \label{fig:sweept15}
\end{figure}

\begin{figure}
\begin{tabular}{cc}
 \hspace*{2.5em}$\mu=0$ & \hspace*{2.5em}$\mu=0.5$ \\
 \begin{minipage}{.25\textwidth}\includegraphics[width = \linewidth]{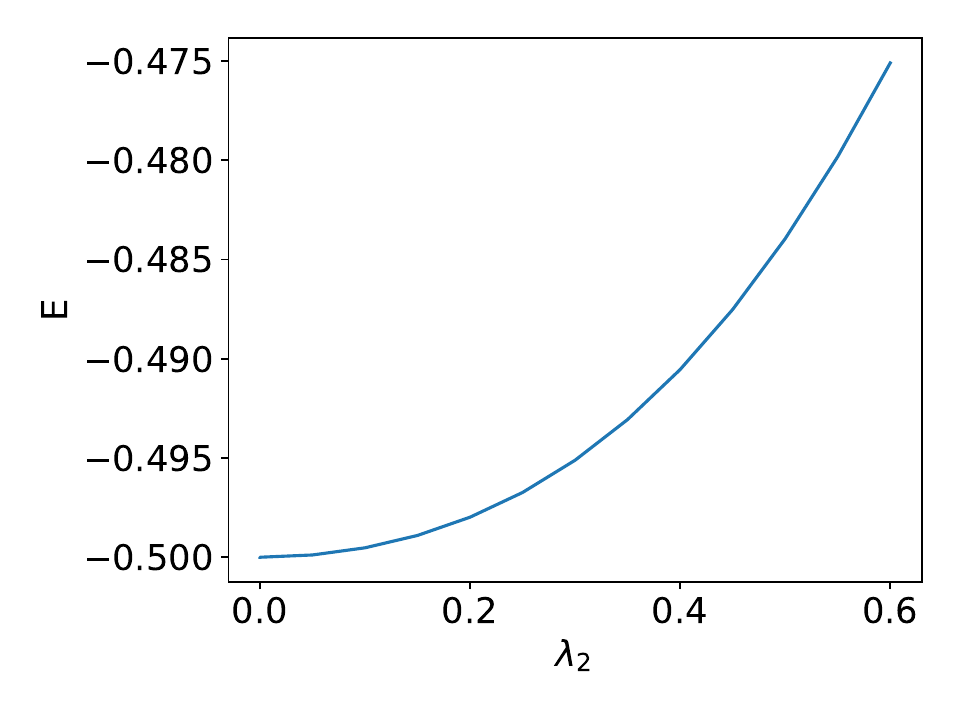}\end{minipage} 
 & \begin{minipage}{.25\textwidth}\includegraphics[width = \linewidth]{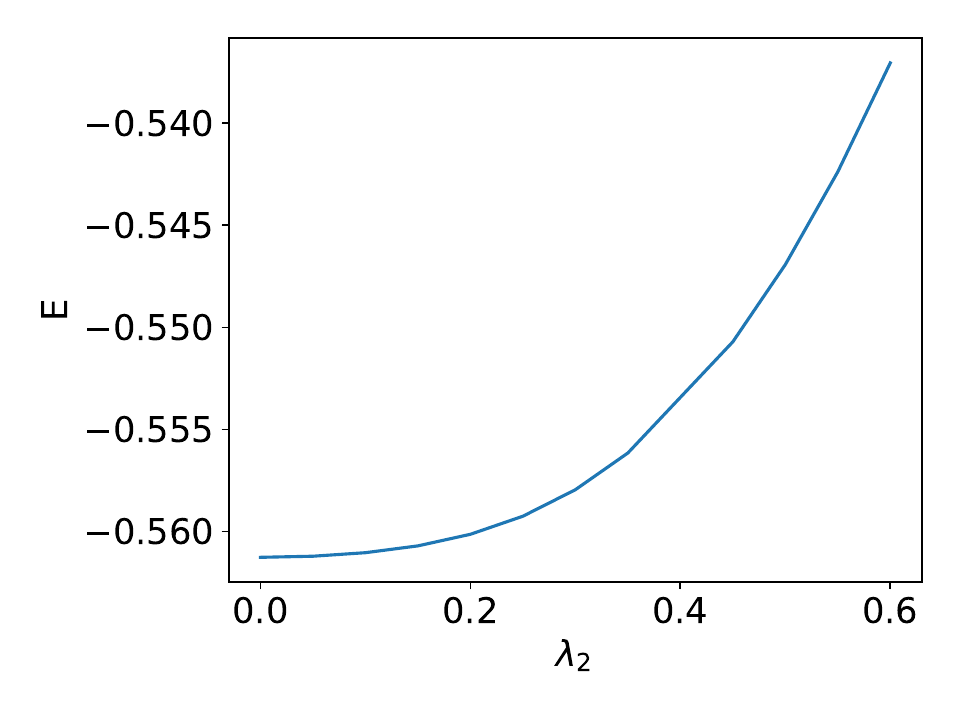}\end{minipage}
\end{tabular}
\begin{tabular}{cc}
 \hspace*{2.5em}$\mu=1.0$ & \hspace*{2.5em}$\mu=1.4$ \\
 \begin{minipage}{.25\textwidth}\includegraphics[width = \linewidth]{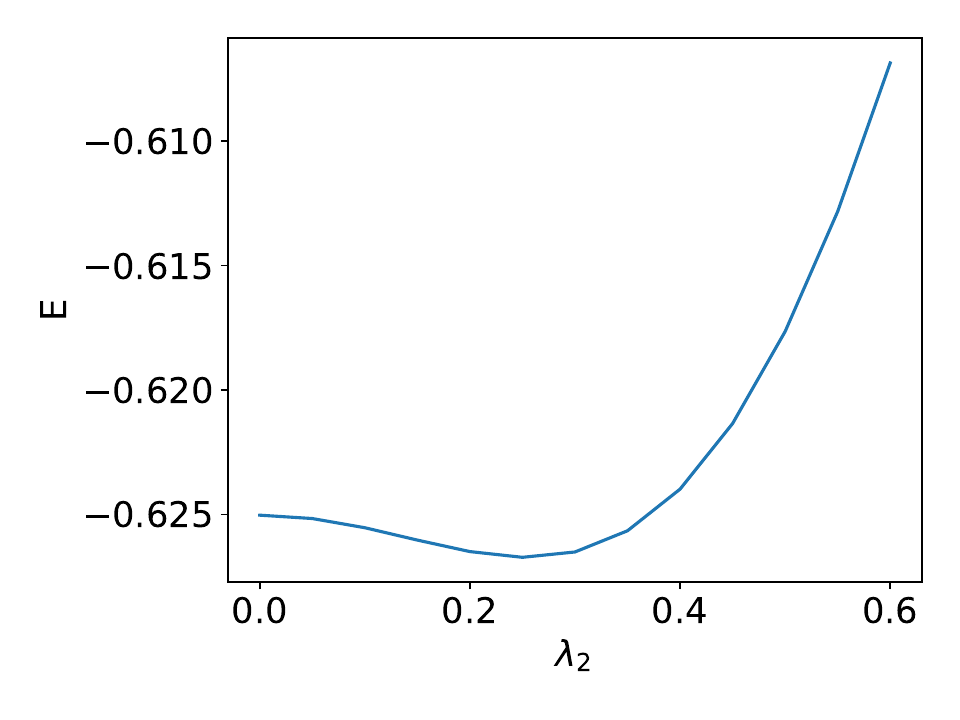}\end{minipage} 
 & \begin{minipage}{.25\textwidth}\includegraphics[width = \linewidth]{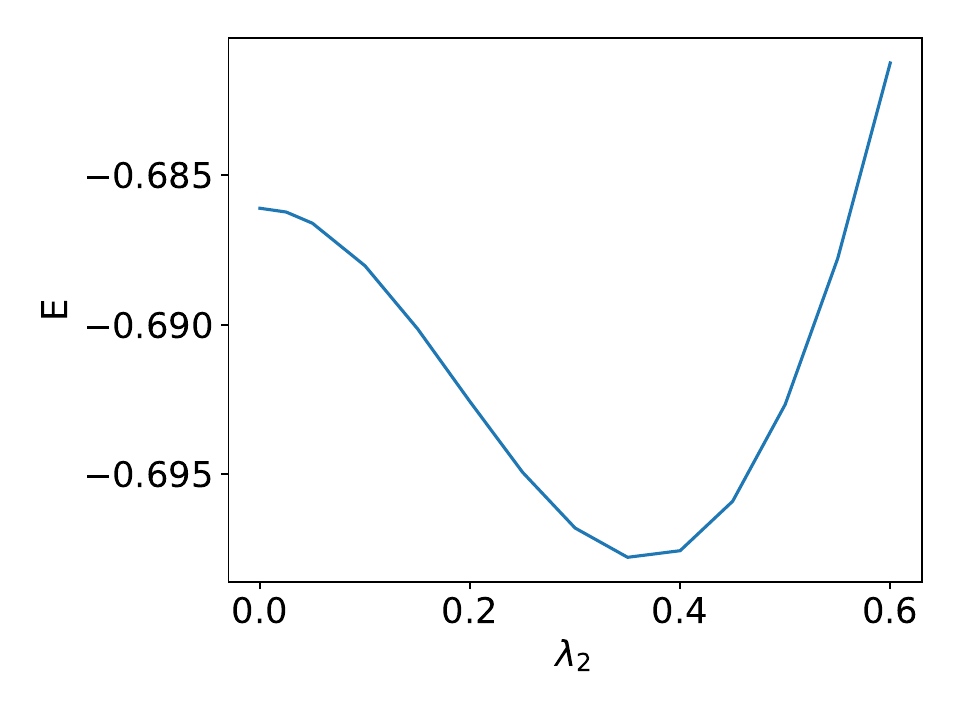}\end{minipage}
\end{tabular}
$\mu=2.0$ 
 \begin{minipage}{.25\textwidth}\includegraphics[width = \linewidth]{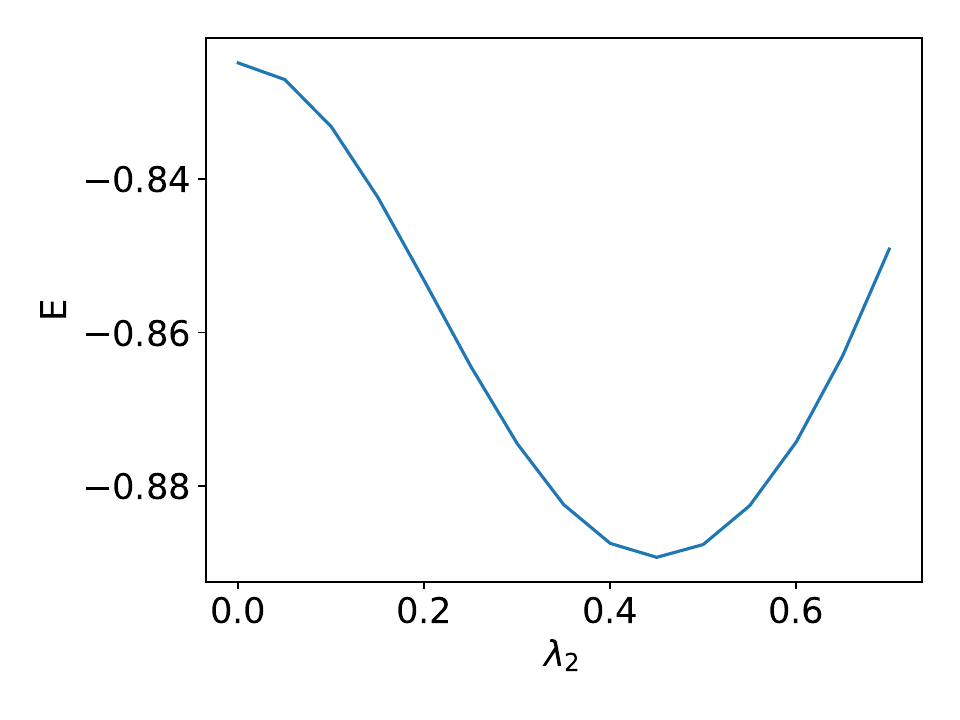}\end{minipage}
\caption{Groundstate energy over $\lambda_2$ for different $\mu$. We see that at a critical $\mu$ a new minimum at a finite $\lambda_2$ is developed, leading to nonzero hole densities and d-wave superconductivity.\label{purehubbarddwave}}
\end{figure}

We apply the above described algorithms and the local unitary transformation to study the pure Hubbard model without $t'$ and investigate first the variation of $t$ in $H_0$, so
\begin{align}
    H_0(\lambda) &= t (1 - \lambda_1) \sum_{\vebm{r},\vebm{\delta}} 
    c^\dag_{\vebm{r}}c_{\vebm{r}+\vebm{\delta}} + \text{h.c.}
    + \sum_{\vebm{r}} \left(- \frac{U}{2} + \mu\right) n_{\vebm{r}},
\end{align}
such that when $\lambda_1=1$, $H_0$ contains no hopping terms and at $\lambda_1=0$ it does. Therefore at $\lambda_1=1$, the vertex $V_\lambda=H_0(\lambda=0)-H_0(\lambda)$ contains the hopping whereas at $\lambda_1=0$ it does not, where $V_\lambda=0$. In the following we set $t=-1$. We consider the $U=8$ strong interaction regime relevant to cuprates. Computing the free energy by minimizing the classical pseudospin groundstate energy we find at $\mu=0$ an optimum at $\lambda_1=1$, see Fig.\,\ref{fig:sweept0}. This optimum exactly replicates the Heisenberg model Eq.\,\eqrefx{Heisenberg}. Even though this is a maximum, this groundstate is chosen by the system, since there the sensitivity (first derivative or second derivative of the energy) to changes in $H_0$ is minimized, see the principle of minimal sensitivity\,\cite{stevenson1984gaussian}. At $\mu=1.5$ (hole doping regime due to the sign convention) this maximum persists, see Fig.\,\ref{fig:sweept15}. So in the following, we fix $\lambda_1=1$. Doing so, we find that at any $0<\mu<U/2$ and $T=0$, the system is unable to fill in holes, see Eq.\,\eqrefx{particledens}, since $\ket{\Psi(\lambda)}$ remains a state fully filled with $c$ fermions. We introduce a variational field into $H_0$ and $V_\lambda$, which at the same time breaks $C_4$ symmetry and is able to introduce holes into the system
\begin{align}
    V_ {\lambda_2} &= \sum_{\posx,\vebm{\delta}}\lambda_2 e^{i(\vebm{Q}\posx+\pi\delta_y)}c^\dag_{\vebm{r}}c^\dag_{\vebm{r}+\vebm{\delta}} + \text{h.c.}
\end{align}
with $\vebm{Q}=(\pi,\pi)$. We find that this field stabilizes d-wave superconductivity. We note that in the pure Hubbard model, we do not only expect to see superconductivity but also for example magnetic stripe orders\,\cite{PhysRevX.10.031016}, which are preferred by the system by $0.01t$ over superconductivity at a doping of $12.5\%$\,\cite{doi:10.1126/science.aam7127}. However, even if a different state is lower in energy, we still expect the system to prefer d-wave superconductivity over pure AFM and therefore to see an optimum appearing at a finite $\lambda_2$, which is exactly what we find, see Fig.\,\ref{purehubbarddwave}. A mexican hat potential develops at a critical $\mu\approx0.7$ and remains at any finite doping. This critical chemical potential is comparable to the one found in\,\cite{metalinsulatorakiyama} for the same system parameters. We suspect that at higher dopings, the approximation of classical spins no longer holds and quantum fluctuations lead to the disappearance of this phase. Nevertheless, the onset of superconductivity is well described by the classical groundstate of a spin Hamiltonian. At $\mu=1.4$, the optimal $\lambda_2=0.37$ corresponds to a doping of $2.4\%$ (the expectation value was computed up to first order in perturbation) and d-wave superconductivity is preferred over pure AFM by an energy of $0.011t$, so $163.4\,\text{K}$ at $t=-1\,\text{eV}$. We find as expected, that a finite chemical potential acts as a magnetic field for the pseudospins, resulting in a $U(1)$ symmetry breaking. The expectation value $|\langle\Delta_{\posx,\posp_x}\rangle|$ is $1.9\cdot 10^{-3}$ in zeroth order on a bond in the $x$ direction, while $\langle\Delta_{\posx,\posp_y}\rangle=-\langle\Delta_{\posx,\posp_x}\rangle$, realizing uniform d-wave superconductivity. We have not yet found an order parameter implementing s-wave or stripe order. We note, that the presented method severely undershoots groundstate energies at higher dopings\,\cite{doi:10.1126/science.aam7127}, which we attribute to the missing energy from the groundstate quantum fluctutations of the pseudospins. 

To test the convergence, the same calculation, which was done on a 16x16 BZ grid for the integration of interaction vertices, was carried out for $\mu=1.4$ on a 32x32 BZ grid. The average error between the 16x16 and the 32x32 calculation is $5\cdot10^{-6}$, meaning that the results are well converged, even at a temperature of $T=10^{-4}$ for the density matrix computations. This signifies, that in all integrations carried out here, no significant singularities have been integrated over. This is the result of modes being integrated out, that are with $U/2$ strongly gapped. This suggests, that the perturbative series described in this paper is controlled and convergent for large $U$ and that higher orders contribute less than the ones we accounted for here.

\begin{figure}
    \centering
    \includegraphics[width = 0.5 \textwidth]{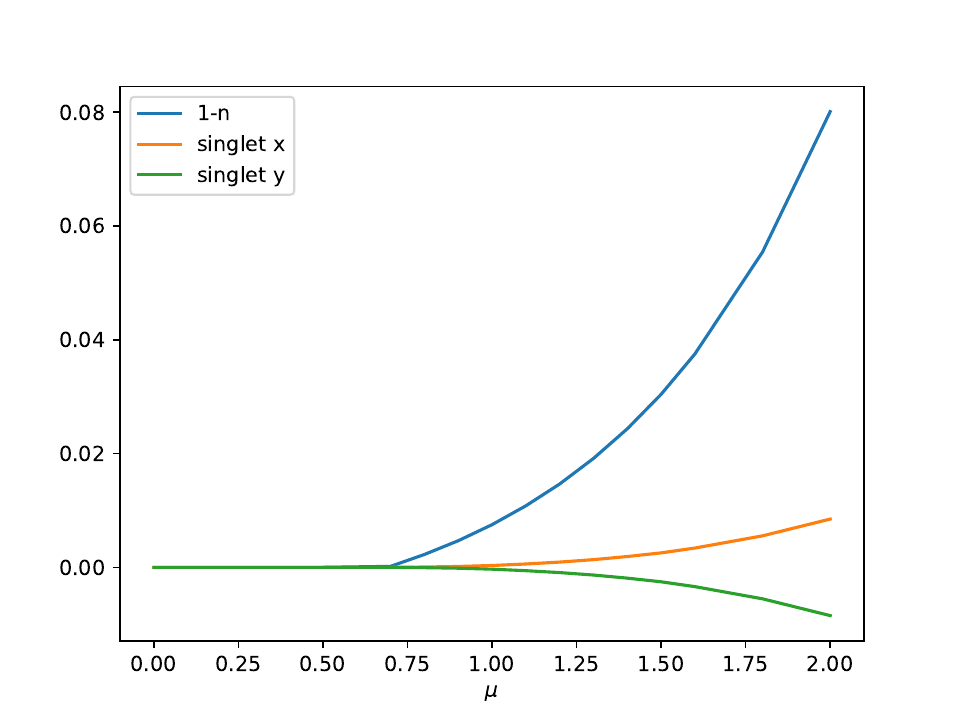}
    \caption{Hole densities $1-n$ and expectation values of $\Delta$ (singlet x/y) over $\mu$ at optimized $\lambda_2$ for $U=8$ in the pure Hubbard model. Since the $U(1)$ phase of both singlets is arbitrary, singlet x has been rotated to be purely real and positive. The rotation has also been applied to singlet y. The imaginary part of singlet y is numerically small. We see a phase transition at $\mu\approx0.7$ to a d-wave superconducting phase. At smaller $\mu$, the hole densities remain at $0$. For every $\mu$, $\lambda_2$ has been optimized to minimize the free energy up to an error of $10^{-4}$.}
    \label{fig:musweep}
\end{figure}

In Fig.\,\ref{fig:musweep}, hole densities and singlet Cooper pair expectation values on nearest neighbor bonds in x and y direction are shown over chemical potential $\mu$, where at every $\mu$ a $\lambda_2$ is found to minimize the free energy up to an error of $10^{-4}$. Hole densities were computed up to first and the singlet expectation values up to zero order in perturbation theory. We find a critical $\mu\approx0.7$ for $U=8$, where at the same time holes start to fill into the system while d-wave superconductivity develops. At smaller $\mu$, the system remains at half-filling.

\begin{figure}
    \centering
    \includegraphics[width = 0.45 \textwidth]{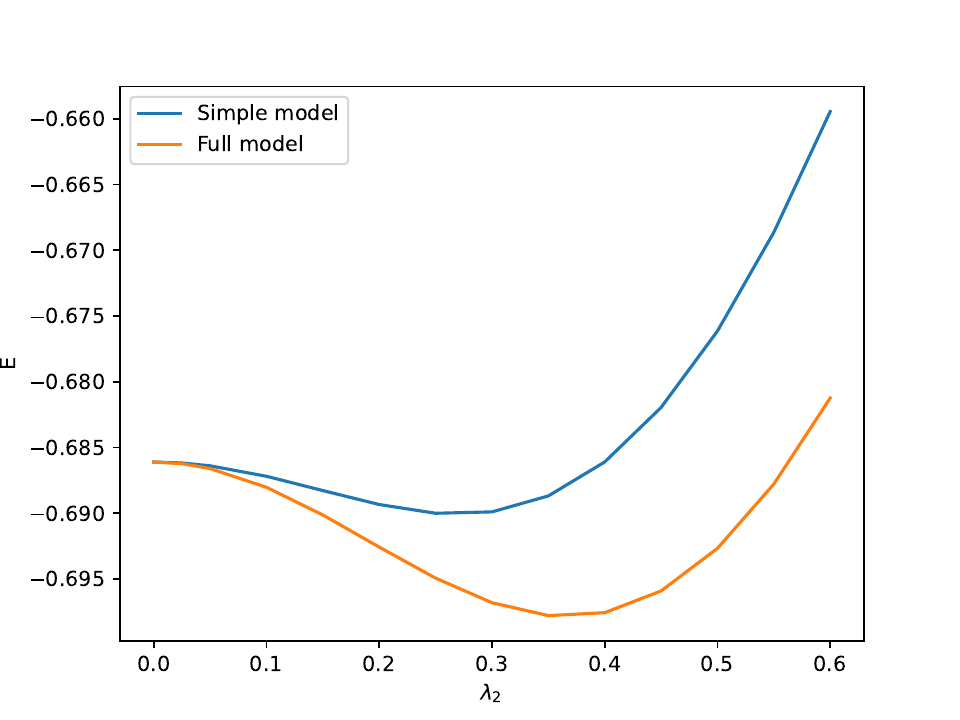}
    \caption{Comparison of groundstate energy over $\lambda_2$ for the full and simplified model at $\mu=1.4$. The terms left out still contribute significantly. Nevertheless, the included terms are enough to observe the development of an energy minimum at a non-zero $\lambda_2$. See Appendix\,\ref{app1} for a description of the simple model.}
    \label{fig:simplemod}
\end{figure}

We have identified the most crucial terms to see the minimum in $\lambda_2$ at a non-zero value and therefore the development of superconductivity. This simplified model is described in Appendix\,\ref{app1}. A comparison of groundstate energies between the full algorithmically build model and this simple model is shown in Fig.\,\ref{fig:simplemod}. We see, that more terms contribute significantly, so this simple model is just a rough approximation of the full one.

\vspace{-1em}
\subsection{Hubbard model at $U=8$ with $t'=-0.3t$}
\begin{figure}
    \centering
    \includegraphics[width = 0.45 \textwidth]{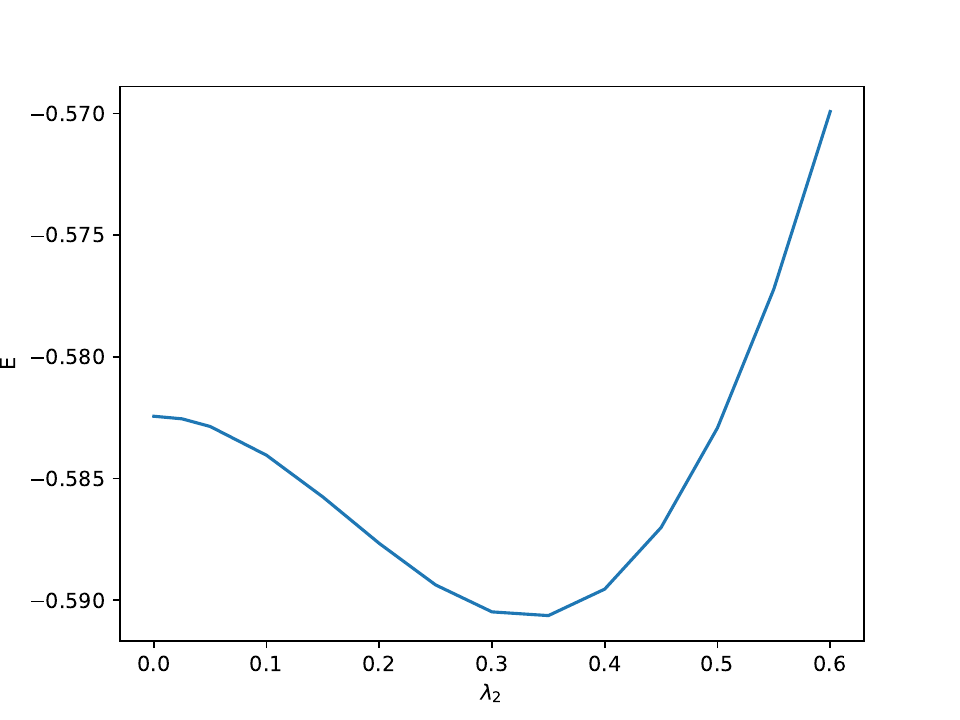}
    \caption{Groundstate energy over order parameter $\lambda_2$ for TBCO/YBCO hopping strengths $t=-1$, $t'=0.3$ and at $U=8$\,\cite{pavarini2001band}. The minimum at a finite $\lambda_2$ remains, signifying the development of superconductivity.}
    \label{fig:YBCOmu14}
\end{figure}
\begin{figure}
    \centering
    \includegraphics[width = 0.5 \textwidth]{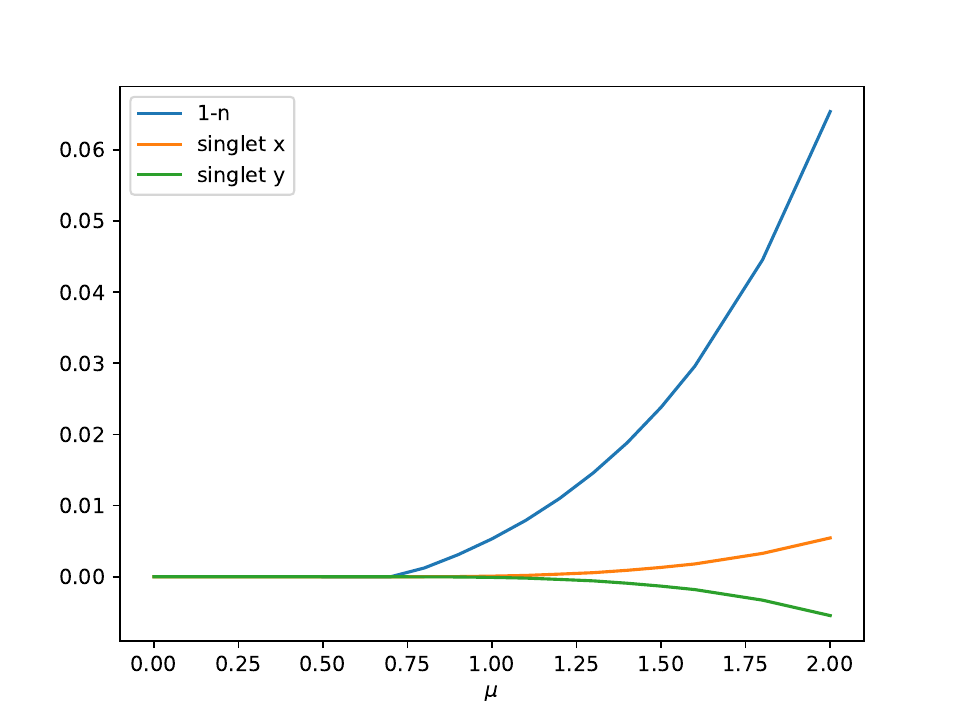}
    \caption{Hole densities $1-n$ and expectation values of $\Delta$ (singlet x/y) over $\mu$ at optimized $\lambda_2$ for $U=8$ in the Hubbard model with $t'=0.3$, at TBCO/YBCO parameters\,\cite{pavarini2001band}. The development of d-wave superconductivity is observed.}
    \label{fig:YBCOholes}
\end{figure}
\begin{figure}
    \centering
    \includegraphics[width = 0.5 \textwidth]{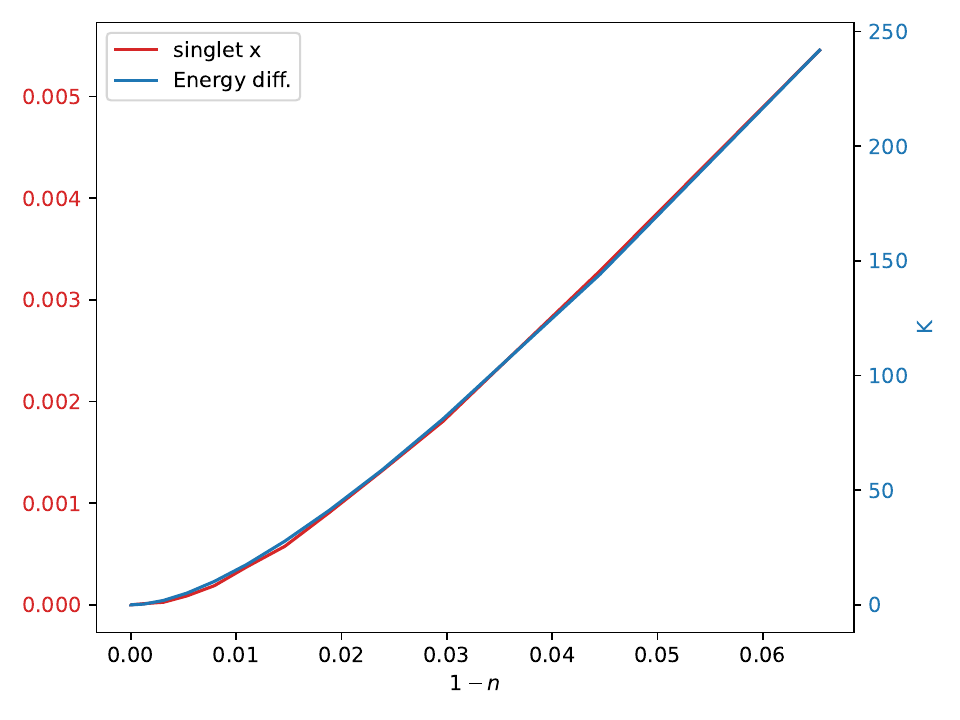}
    \caption{Energy difference between the groundstate energy at optimal $\lambda_2$ and $\lambda_2=0$ in Kelvin (for $t=0.43\,\text{eV}$\,\cite{pavarini2001band}) and singlet pairing expectation values over hole doping for TBCO/YBCO parameters.}
    \label{fig:YBCOdope}
\end{figure}
We apply the same methods described in the previous sections to TBCO/YBCO hopping parameters with $t'=-0.3t$ being the next nearest neighbor hopping amplitude\,\cite{pavarini2001band}. $U$ is set to $8$. The $\lambda_2$ groundstate energy potential is shown in Fig.\,\ref{fig:YBCOmu14}. The minimum in the free energy remains at a finite value of $\lambda_2=0.33$. %This corresponds to a doping of $3\%$. The expectation value of $\Delta$ becomes $1\cdot10^{-3}$. The energy difference between the maximum at $\lambda_2=0$ and the minimum in energy $\lambda_2=0.35$ is $0.0082t$. With $t=-0.43\,\text{eV}$, superconductivity is preferred by a temperature of $40.9\,\text{K}$.
In Fig.\,\ref{fig:YBCOholes}, hole densities and singlet pairing expectation values are shown. The latter correspond to the development of d-wave superconductivity. 
Fig.\,\ref{fig:YBCOdope} shows the singlet expectation value with $t'=-0.3t$ and the groundstate energy difference between optimal $\lambda_2$ and $\lambda_2=0$ in Kelvin at $t=0.43\,\text{eV}$\,\cite{pavarini2001band}. We see that this newly developed energy scale is on the order of $T_c$ in these compounds and that the energy difference scales proportionally with the superconducting order parameter. Energy differences are likely overestimated as we expect that quantum fluctuations in the pseudospins weaken the states found at $\lambda_2>0$.

\section{Discussion}

In this paper, we propose a new theoretical description of high temperature superconductivity, based on the application of a unitary transformation to the Hubbard model, which maps the Hubbard interaction to a single particle term. This unitary transformation can be applied exactly. The resulting Hamiltonian contains, unlike the t-J model, unconstrained fermions and can be transformed into one with a single fermion species coupled to hardcore bosons/pseudospins, which suggests that this transformation implements a spin charge separation in any dimension, where the fermions carry charge and the pseudospins spin information. The Hamiltonian is solved by variational second order perturbation theory (VPT), resulting in an effective pseudospin Hamiltonian. The problem of fermions coupled to spins generating an effective spin-spin interaction is reminiscent of the RKKY mechanism\,\cite{PhysRev.96.99}. The breakdown of the phases found in this paper might be equivalent to a Kondo-RKKY transition. To compute the groundstate energy, the spins are replaced by classical spins and their classical energy is minimized. We find that in zero order, the introduction of a chemical potential $\mu$ acts as a magnetic field for the pseudospins, breaking $U(1)$ symmetry. Mapping back to the original system, this $U(1)$ symmetry breaking corresponds to the $U(1)$ symmetry breaking found in a superconducting state, assuming the superconducting state is the one minimizing the free energy. An order parameter is proposed which at the same time fills in holes into the system and breaks $C_4$ symmetry. The free energy is minimized at a finite value of this order parameter, resulting in a finite expectation value of the d-wave superconductor pairing operator. This perturbative approach is possible, because the fermionic modes that are integrated out are strongly gapped with a gap of $U/2$. We therefore encourage further studies of Hamiltonians\,\eqrefx{Hp} or\,\eqrefx{Hf} with non-perturbative calculations (for example DMFT\,\cite{PhysRevLett.62.324,PhysRevB.45.6479,RevModPhys.68.13} or DMRG\,\cite{PhysRevLett.69.2863}), which should even increase the accuracy of the presented results. Future work also involves exploring similar unitary transformations like\,\eqrefx{U}, for example ones that map a whole cluster of sites to single particle terms. It can also be interesting to vary different and more variational parameters than just one in VPT, like the chemical potential. It is also possible to connect auxiliary bath sites with VPT. Exploring the influence of quantum fluctuations around the classical spin groundstate found in this paper should also give more insights. This is especially important for studying magnetism in the models found in this paper, since classical spins can not describe paramagnetic states at $T=0$, which is why we did not explore magnetism here. In light of numerous potential paths for future work, this paper already suggests that the emergence of high-$T_c$ superconductivity is governed by the ground state of a classical spin system, showcasing promising directions for further research.

\section*{Acknowledgements}
I am thankful for Robin Scholle, Lukas Debbeler, Steffen Bollmann, Silvia Neri, Pietro Maria Bonetti, Elio König, Laura Classen, Nikolaos Parthenios, Sida Tian, Raffaele Mazzilli, Henrik Müller-Groeling, Paulo Forni, Janika Reichstetter, Andreas Schnyder and Walter Metzner for the many enlightening discussions. I would also like to thank Thomas Schäfer, Andreas Schnyder and Walter Metzner for reviewing this paper. Kirill Alpin is funded by the Deutsche Forschungsgemeinschaft (DFG, German Research Foundation) – TRR 360 – 492547816.

\vspace{-1em}
\appendix
\renewcommand{\appendixname}{APPENDIX}
\section{Simplest effective pseudospin model\label{app1}}
\vspace{-1em}
We have identified the most crucial terms of the effective pseudospin Hamiltonian in second order perturbation theory to observe the development of d-wave superconductivity. The spin Hamiltonian takes on this form
\begin{align}
    H_S &= \bra{\Psi(\lambda)}H_0(\lambda=0)\ket{\Psi(\lambda)} \\
    &- \bra{\Psi(\lambda=0)}H_0(\lambda=0)\ket{\Psi(\lambda=0)}\label{offset}\\
    &+2 \mu \sum_\posx \left(\frac{1}{2} - S^z_\posx\right) (1 - \langle n_\posx\rangle) \\
    &+ X_1 + X_2 t^2 \sum_{\vebm{r},\vebm{\delta}} (1 - 4 \vebm{S}_\posx \cdot \vebm{S}_\posp) \label{heisterm} \\
    &+(- 2 X_3 \mu + 4 X_4 \mu^2)  \sum_\posx \left(\frac{1}{2} - S^z_\posx\right) \\
    &+ 4 X_5 \mu^2 \sum_{\vebm{r},\vebm{\delta}} \left(\frac{1}{2} - S^z_\posx\right)\left(\frac{1}{2} - S^z_\posp\right) \\
    &+ X_6 t^2 \sum_{\vebm{r}_1,\vebm{r}_2,\vebm{r}_3} W_{\vebm{r}_1,\vebm{r}_2,\vebm{r}_3}
\end{align}
with
\begin{align}
    X_1 &= \bra{\Psi(\lambda)} V_l \frac{1}{E_0(\lambda) - H_0(\lambda)} V_l \ket{\Psi(\lambda)} \\
    X_2 &= \bra{\Psi(\lambda)} c^\dag_{\vebm{r}}c^\dag_{\vebm{r}+\vebm{\delta}} \frac{1}{E_0(\lambda) - H_0(\lambda)} c_{\vebm{r}+\vebm{\delta}}c_{\vebm{r}} \ket{\Psi(\lambda)} \\
    X_3 &= \bra{\Psi(\lambda)} V_l \frac{1}{E_0(\lambda) - H_0(\lambda)} n_\posx \ket{\Psi(\lambda)} + \text{h.c.} \\
    X_4 &= \bra{\Psi(\lambda)} n_\posx \frac{1}{E_0(\lambda) - H_0(\lambda)} n_\posx \ket{\Psi(\lambda)} \\
    X_5 &= \bra{\Psi(\lambda)} n_\posx \frac{1}{E_0(\lambda) - H_0(\lambda)} n_{\posp} \ket{\Psi(\lambda)} \\
    X_6 &= \bra{\Psi(\lambda)} c^\dag_{\vebm{r}_1}c^\dag_{\vebm{r}_2} \frac{1}{E_0(\lambda) - H_0(\lambda)} c_{\vebm{r}_2}c_{\vebm{r}_3} \ket{\Psi(\lambda)}
\end{align}
and
\begin{align}
    W_{\vebm{r}_1,\vebm{r}_2,\vebm{r}_3} &= K_{\vebm{r}_1,\vebm{r}_2} K^\dag_{\vebm{r}_3,\vebm{r}_2}\\
    &= -\frac{1}{4} + \vebm{S}_{\vebm{r}_1}\cdot\vebm{S}_{\vebm{r}_2}
    + \vebm{S}_{\vebm{r}_2}\cdot\vebm{S}_{\vebm{r}_3} \nonumber\\
    &- \vebm{S}_{\vebm{r}_1}\cdot\vebm{S}_{\vebm{r}_3} - 2 i \sum_{nmp} \epsilon_{nmp} S^n_{\vebm{r}_3} S^m_{\vebm{r}_2} S^p_{\vebm{r}_1}.
\end{align}
$\epsilon_{nmp}$ is the Levi-Civita tensor and the sum $\sum_{\vebm{r}_1,\vebm{r}_2,\vebm{r}_3}$ runs over all nearest-neighbor L shaped 3 spin clusters, where $\vebm{r}_1$ and $\vebm{r}_3$ are the furthest apart. All $X_n$ are isotropic in $\vebm{r}$ and $\vebm{\delta}$, so it does not matter where they are evaluated. The term\,\eqrefx{offset} is not necessary and is applied to measure relative energies.
You arrive at the Heisenberg term\,\eqrefx{heisterm} by $K_{\posx,\vebm{\delta}}K^\dag_{\posx,\vebm{\delta}}=1 - 4 \vebm{S}_\posx \cdot \vebm{S}_\posp$.
\bibliographystyle{apsrev4-1}
\bibliography{bib.bib}

\end{document}